\documentclass[final,authoryear,5p]{elsarticle}

\usepackage{epsfig}

\usepackage{amssymb}

\usepackage[ps2pdf,%
a4paper=true,%
breaklinks=true,%
colorlinks=true,%
pdfauthor={First Author et al.},%
pdftitle={Template for manuscripts in Advances in Space Research}%
]{hyperref}

\journal{Advances in Space Research}

\begin{document}

\begin{frontmatter}



\title{The STARK-B database as a resource for \textquotedblleft STARK"  widths and shifts data: State of advancement and program of development \tnoteref{footnote1}}
\tnotetext[footnote1]{in press, available on line at www.sciencedirect.com: J. Adv. Space Res. (2013), http://dx.doi.org/10.1016/j.asr.2013.08.015}


\author[label1]{Sylvie Sahal-Br\'echot}
\ead{sylvie.sahal-brechot@obspm.fr}
\address[label1]{Laboratoire d'\'{E}tude du Rayonnement et de la
Mati\`{e}re en Astrophysique,Observatoire de Paris,\\
UMR CNRS 8112, UPMC, 5 Place Jules Janssen, 92195 Meudon Cedex, France.}


\author[label1,label2]{Milan S. Dimitrijevi\'c}
\ead{mdimitrijevic@aob.bg.ac.rs}
\address[label2]{Astronomical Observatory, Volgina 7, 11060 Belgrade,
Serbia.}

\author[label1]{Nicolas Moreau}
\ead{nicolas.moreau@obspm.fr@email.addresses}

\author[label3]{Nabil Ben Nessib}
\address[label3]{Department of Physics and Astronomy, College of Science, King Saud University. Riyadh 11451, Saudi Arabia}
\ead{nbennessib@ksu.edu.sa}

\begin{abstract}
\textquotedblleft Stark" broadening theories and calculations have been extensively developed for about 50 years and can now be applied to many needs, especially for accurate spectroscopic diagnostics and modeling. This requires the knowledge of numerous collisional line profiles. Nowadays, the access to such data via an online database becomes essential.
STARK-B  is a collaborative project between the Astronomical Observatory of Belgrade and the Laboratoire d'\'Etude du Rayonnement et de la mati\`ere en Astrophysique (LERMA). It is a database of calculated widths and shifts of isolated lines of atoms and ions due to electron and ion collisions (impacts). It is devoted to modeling and spectroscopic diagnostics of stellar atmospheres and envelopes, laboratory plasmas, laser equipments and technological plasmas. Hence, the domain of temperatures and densities covered by the tables is wide and depends on the ionization degree of the considered ion. STARK-B has been fully opened since September 2008 and is in free access. 

The first stage of development was ended in autumn 2012, since  all the existing data calculated with the impact semiclassical-perturbation method and code by Sahal-Br\'echot, Dimitrijevi\'c and coworkers have now been implemented. 
We are now beginning the second stage of the development of STARK-B. 
The state of advancement of the database and our program of development are presented here, together with its context within VAMDC. VAMDC (Virtual Atomic and Molecular Data Center) is an international consortium which  has built a secure, documented, flexible interoperable platform e-science  permitting an automated exchange of atomic and molecular data.
\end{abstract}


\end{frontmatter}

\parindent=0.5 cm

\section{Introduction}

Broadening of spectral lines due to perturbation of emitting or absorbing atoms, ions, or molecules, by their interactions with the surrounding particles in a plasma or gas is called pressure broadening. If perturbers are charged particles, electrons or ions, such broadening mechanism is called Stark broadening. Data on Stark broadening are of interest for plasma diagnostic and modeling, as well as for spectra analysis and synthesis in many domains for research of different kinds of plasma in laboratory, inertial fusion investigations, for lasers and laser produced plasmas and especially in astrophysics, since thanks to space missions, large space observatories like Hubble, Chandra, Spitzer, Lyman, and to ground-based telescopes of the ten meters class, spectra with very high resolution could be obtained from X to radio wavelength ranges.

In stellar astronomy, Stark broadening data are of particular importance for white dwarfs, where Stark broadening is the dominant collisional line broadening mechanism
\citep{Pop99b,Tan03,Sim06,Ham08,Sim09,Dim11,Duf11,Lar12}. Such data are also of significance for interpretation, analysis and synthesis of A and B type star spectra (see e.g. \citet{Lan88}, \citet{Pop01a,Pop01b,Pop99a,Pop99b}, \citet{Tan03},
\citet{Sim05}, \citet{Sah10}. Modern codes for stellar atmosphere modeling, like e.g. PHOENIX \citep{Bar98, Hau99, Sho99} require the knowledge of atomic data for as much as possible transitions, especially for trace elements, so that the access to such atomic data, including Stark broadening ones, via online databases becomes very important.

The database STARK-B \citep{Sah13}, which contains Stark broadening parameters, namely calculated line widths and shifts of isolated lines of atoms and ions due to collisions with electrons, protons and different ions, first of all ionized helium as the most important after electrons and protons for stellar atmospheres, is not useful only for modeling and spectroscopic diagnostics of stellar atmospheres and envelopes, but also for laboratory plasmas, inertial fusion plasma, laser equipments and technological plasmas. This is a common project where participants are from the Paris Observatory and the Astronomical Observatory of Belgrade. The database STARK-B has been opened online since September 2008, and is in free access. The objective of the first phase of database development, was to implement all the existing data calculated with the impact semiclassical-perturbation method and code by Sahal-Br\'echot, Dimitrijevi\'c and coworkers.  We are now working on the second stage and the state of advancement of the database STARK-B will be presented here, as well as our program of development. We note also that STARK-B enters in Virtual Atomic and Molecular Data Center - VAMDC \citep{Dub10, Rix11}, a secure, documented, flexible, interoperable platform based on e-science,  which permits an optimized search and exchange of atomic and molecular data. The relation between STARK-B database and VAMDC is discussed in \citet{Sah12}.

\section{STARK-B database}

 Within the impact-complete collision-isolated line approximation, the profile of an isolated spectral line is  the Lorentz one, characterized by a width at half maximum of intensity and a shift, so called Stark broadening parameters, depending on  temperature T and density N of perturbers. For not too high densities, due to the impact approximation, widths and shifts are linearly proportional to the density of perturbers. But at high densities, due to the Debye screening effect which decreases the width and the shift, and which for such conditions can become significant, the dependence of Stark broadening parameters as a function of perturber density may decline from the linear one. This is taken into account in majority of our calculations, so that for many of atoms and ions the Stark broadening parameters for a grid of perturber densities are implemented in STARK-B.

In the first stage of STARK-B development, we implemented in STARK-B Stark broadening parameters determined with the semiclassical-perturbation method (SCP) developed by \citet{Sah69a, Sah69b} and updated in further papers by \citet{Sah74}) for more complex atoms, by \citet{Fleu77} for inclusion of Feshbach resonances in elastic cross-sections for the line widths for ion emitters, and by \citet{mah08} for the particular cases of very complex atoms. The numerical codes have been also updated by  \citet{dimi84} and in further papers. The accuracy is about 20\% for the widths of simpler spectra but is worse for the shifts and very complex spectra, particularly when configuration mixing is present in description of energy levels. The results of Stark broadening parameters determination performed by Dimitrijevi\'c, Sahal-Br\'echot, and co-workers using the semiclassical-perturbation method are contained in more than 130 publications and now have been implemented in the STARK-B database.

STARK-B is devoted for modeling and spectroscopic diagnostics of various plasmas in astrophysics, laboratory physics, technology and other topics, so that the range of temperatures and densities in the tables is wide and vary with the  ionization degree of the considered ion. The temperatures in tables vary from several thousands Kelvins for neutral atoms to several millions for highly charged ions. Also the perturber densities vary from 10$^{12}$ cm$^{-3}$ to several 10$^{22}$cm$^{-3}$. The presentation of data, in particular the definition of configurations, terms and atomic energy levels is in accordance with the VAMDC standards, in order to allow interoperability with other atomic databases included in the Virtual Atomic and Molecular Data Center. We note also that the wavelengths in STARK-B are usually calculated from the energy levels that are used as input data for the determination of Stark broadening parameters.  Consequently, the tabulated wavelengths are most often different from the measured ones, especially if the used energy levels are theoretically calculated. For the identification of the lines the configurations, terms and levels are included in tables as well as the multiplet number from NIST database \citep{Ral08}.

If one opens the homepage, proposed menus are  \textquotedblleft Introduction",  \textquotedblleft Data Description",  \textquotedblleft Access to the Data", \textquotedblleft Updates" and  \textquotedblleft Contact". In  \textquotedblleft Introduction" methods of Stark broadening parameter calculations and different approximations which are adopted are briefly explained.  \textquotedblleft Data Description" describes the data which are tabulated.  \textquotedblleft Access to the Data" provides a graphical interface enabling to the visitor to click on the needed element in the Mendeleev periodic table and then on the corresponding ionization degree. For elements in yellow cells, which symbols are additionally enhanced by boldface, Stark broadening data exist in the database, and for other cells, which have the color a little bit lighter than  the background, there is no data. After choosing the needed ionization stage, the user with several clicks chooses the colliding perturber(s), the perturber density, the transition(s) by quantum numbers and the plasma temperature(s). A query by domain of wavelengths instead by transitions is also possible. Then a table containing the Stark full widths at half maximum of intensity and shifts is generated. In our tables,  a positive shift is towards the red and a negative one towards the blue. At the beginning, before the table an instruction how to cite the use of STARK-B, and bibliographic references for the data in the table below are given and linked to the publications via the SAO/NASA ADS Physics Abstract Service (http://www.adsabs.harvard.edu/) and/or within DOI, if available. The widths and shifts data can be downloaded as an ASCII table or in format adapted for Virtual Observatories - VOTable format (XML format).

Actually (1st of July 2013) in the STARK-B are Stark broadening parameters for 79 transitions of He, 61 Li, 29 Li II, 19 Be, 30 Be II, 27 Be III, 1 B II, 12 B III, 148 C II, 1 C III, 90 C IV, 25 C V, 1 N, 7 N II, 2 N III, 1 N IV, 30 N V, 4 O I, 12 O II, 5 O III, 5 O IV, 19 O V, 30 O VI, 14 O VII, 8 F I, 5 F II, 5 F III, 2 F V, 2 F VI, 10 F VII, 25 Ne I, 22 Ne II, 5 Ne III, 2 Ne IV, 26 Ne V, 20 Ne VIII, 62 Na, 8 Na IX, 57 Na X, 270 Mg, 66 Mg II, 18 Mg XI, 25 Al, 23 Al III, 7 Al XI, 3 Si, 19 Si II, 39 Si IV, 16 Si V, 15 Si VI, 4 Si XI, 9 Si XII, 61 Si XIII, 114 P IV, 51 P V, 6 S III, 1 S IV, 34 S V, 21 S VI,2 Cl, 10 Cl VII, 18 Ar, 2 Ar II, 9 Ar VIII,51 K, 4 K VIII, 30 K IX, 189 Ca, 28 Ca II, 8 Ca V, 4 Ca IX, 48 Ca X, 10 Sc III, 4 Sc X, 10 Sc 11,10 Ti IV, 4 Ti XI, 27 Ti XII,26 V V, 33 V XIII, 9 Cr I, 7 Cr II, 6 Mn II, 3 Fe II, 2 Ni II, 9 Cu I, 32 Zn, 18 Ga, 11 Ge, 11 Kr, 1 Kr II, 6 Kr VIII, 24 Rb, 33 Sr, 32 Y III, 3 Pd, 48 Ag, 70 Cd, 1 Cd II, 18 In II, 20 In III, 4 Te, 4 I, 14 Ba, 64 Ba II, 6 Au, 7 Hg II, 2 Tl III and 2 Pb IV.

We will continue to implement the new results and under the menu  \textquotedblleft Updates" is the description of newly added data with the date of importation. Also all updates with the date of the first importation and the importation of revised data are noted. Moreover, for further enquiries or user support, there is the menu  \textquotedblleft Contact" with the possibility to send an e-mail with questions to the corresponding persons.

\section{Fitting formulae as functions of temperature}

The first step of the stage two of STARK-B development was the implementation of possibility to fit the tabulated data with temperature. The theory of Stark broadening shows that the dependence of line widths with temperature is proportional to $T^{-1/2}$ at low temperature and to $\mathrm{log}(T)/T^{1/2}$ at high temperatures. This was checked many times, for example in \citet{Ela09}. However, in astrophysics, especially for the modeling of stellar atmospheres, this is not sufficient, since fitting formulae and coefficients as functions of temperature for the whole line are needed, since such fitting coefficients are easier to be imported to the computer codes for stellar atmosphere modeling than  data with tabulated  widths and shifts for a set of temperatures.

Consequently, in order to enable the better and more adequate use of STARK-B for stellar atmospheres modeling,  we have derived \citep{Sah11} a simple and accurate fitting formula based on a least-square method, which is logarithmic and represented by a second degree polynomial:
\begin{equation}
\begin{array}{l}
 \log (w) = a_0  + a_1 \log (T) + a_2 \left( {\log (T)}\right)^2 , \\
 d/w = b_0  + b_1 \log (T) + b_2 \left( {\log (T)}\right)^2 . \\
 \end{array}
\end{equation}

One can be noted that \citet{Dim07} proposed the fitting formula $w= C + A T^{B}$, but the present one is more accurate, due to the second degree term of the expansion. It should be noted also that none of them have a real physical sense.

Now in STARK-B, for each table with widths and shifts is added a complementary table with coefficients $a_0$, $a_1$, $a_2$ and $b_0$, $b_1$, $b_2$ for the corresponding fitting with the temperature using Eq. (1), so that it is easier to include the corresponding Stark broadening parameters in computer codes for stellar atmospheres modeling. We plan also to develop the fitting formulae as functions of perturber densities in order to make easier the use of data on high densities needed for white dwarf atmospheres and sub photospheric layers modeling.

\section{The further development of STARK-B, stage 2}

As the next step in STARK-B development is planned the implementation of Stark broadening data obtained with the Modified semiempirical method \citep{Dim80, Dim86, Dim01}. We use  this approach for emitters where
atomic data are not sufficiently complete to perform an adequate
semiclassical perturbation calculation.  Stark line widths and in
some cases also shifts of the following emitters spectral lines
were calculated up to now:

Ag II, Al III, Al V, Ar II, Ar III, Ar IV, As II, As III, Au II, B III, B IV, Ba II, Be III, Bi II, Bi III, Br II, C III, C IV, C V, Ca II, Cd II, Cl III, Cl IV,  Cl VI, Co II, Cu III, Cu IV, Eu II, Eu III, F III, F V, F VI, Fe II, Ga II, Ga III, Ge III, Ge IV, I II, Kr II, Kr III, La II, La III, Mg II, Mg III, Mg IV, Mn II, N II, N III, N IV, N VI, Na III, Na VI, Nd II, Ne III, Ne IV, Ne V, Ne VI, O II, O III, O IV, O V, P III, P IV, P VI, Pt II, Ra II, S II, S III, S IV, Sb II, Sc II, Se III, Si II, Si III, Si IV, Si V, Si VI,
Sn III, Sr II, Sr III, Ti II, Ti III, V II, V III, V IV, Xe II, Y II, Zn II, Zn III, and Zr II.

Another future step is the implementation of our quantum data in STARK-B. Also, our plans concerns the development of additional fittings along principal quantum number for a given multiplet, charge of the ion collider along isoelectronic sequences, of the radiating ion, homologous ions… in order to enable to estimate by interpolation and extrapolation the data that are missing in STARK-B database.

\section*{Acknowledgments}
The support of Ministry of Education, Science and Technological Development
of Republic of Serbia through projects 176002 and III44022 is acknowledged. A part of this work has been supported by VAMDC. VAMDC is funded under the Combination of Collaborative Projects and Coordination and Support Actions  Funding Scheme of The Seventh Framework Program. Call topic: INFRA-2008-1.2.2 Scientific Data Infrastructure. Grant Agreement number: 239108. This work has also been supported by the  cooperation agreements between Tunisia (DGRS) and France (CNRS) (project code 09/R 13.03, No.22637), and by the Programme National de Physique Stellaire (INSU-CNRS). The support of the French LABEX Plas@Par (UPMC, PRES Sorbonne Universties) is also acknowledged.

\end{document}